# ELECTRICAL CONDUCTANCE OF NANOFLUIDIC SYSTEMS SUBJECTED TO ASYMMETRIC CONCENTRATIONS

Oren Lavi and Yoav Green*

Department of Mechanical Engineering, Ben-Gurion University of the Negev, Beer-Sheva 8410501, Israel

A nanochannel subjected to both a potential and concentration gradient has an asymmetric current-voltage, $I-V$, response with three primary characteristics: the ohmic conductance, $G_{\text{Ohmic}} = I/V$, the current at zero voltage, $I_{V=0}$, and the voltage at zero current, $V_{I=0}$. To date, there is no known self-consistent theory for these characteristics subject to an arbitrary concentration gradient. Here, we present simple expressions for each of these characteristics that have been derived self-consistently. Our findings provide insights into the underlying physics of nanofluidics systems used for water desalination and energy harvesting.

**Introduction.** Elucidating the fundamental physics of ion transport across ion-selective materials is a crucial and required step in revolutionizing the established processes of electrodialysis (ED) and reverse-electrodialysis (RED) used for water desalination and energy harvesting [1–3]. In these processes, ion-selective materials are placed between two large reservoirs of an electrolyte of asymmetric salt concentrations [**Figure 1**(a)]. Since the ion-selective material has an inherent surface charge, a symmetry-breaking process ensues. Here, counterions (ions that are oppositely charged to the surface charge density) are freely transported across the material, while coion (ions of the same sign) transport is restricted. The selective capability of these materials lies at the heart of ED and RED [4,5] processes and is commonly termed permselectivity.

For all practical purposes, any material can be permselective. All that is required is that the surface be charged, even slightly, the dimensions are adequately small, and that the electrolyte concentration is sufficiently low. Once these conditions are satisfied, all materials (carbon nanotubes [6–8] and carbon nanotubes porins [9,10], boron nitride [11], silicon nanochannels and nanopores [12–16], graphene and graphene-oxide membranes [17,18], conducting hydrogels [19], wood-based cellulose [20,21], biological pores [22], single layer $MoS_2$ [23–25], and many additional materials [26–28], see the recent review for a detailed list [5]) can be colloquially termed permselective. In the remainder, we will address the simplest form of a permselective system – this is the nanochannel system, which has a very well-defined geometry.

Importantly, all materials, regardless of their macroscopic and microscopic structure, can be electrically characterized via their current-voltage response $I-V$. If the concentrations at the two ends of the system are symmetric, i.e., equal, the $I-V$ curve goes through the origin (red line in **Figure 2**), where the ratio gives the electrical conductance, $G = I/V$ (inset in **Figure 2**). However, if the concentrations at two ends of the system are asymmetric, i.e., not equal, the $I-V$ curve is shifted (brown line in **Figure 2**) and is also characterized by the two intercept points, which are the current at zero voltage, $I_{V=0}$, (also known as short-circuit current, osmotic current [23,24], and streaming current [11]) and the voltage at zero current, $V_{I=0}$ (also known as open circuit voltage, osmotic voltage [23,24], and reversal voltage [7]).

To date, the conductance $G$ is not known for asymmetric systems, $V_{I=0}$ is known only for ideally selective systems, and $I_{V=0}$ can only be determined experimentally. In this work, we provide expressions for each of these quantities. The expressions have been derived from the Poisson-Nernst-Planck equations – the details of which are given in the Supplementary Information [30]. We then compare our theoretical model to non-approximated numerical simulations, where one finds that the correspondence is remarkable. See Ref. [30] for details regarding numerical simulations.

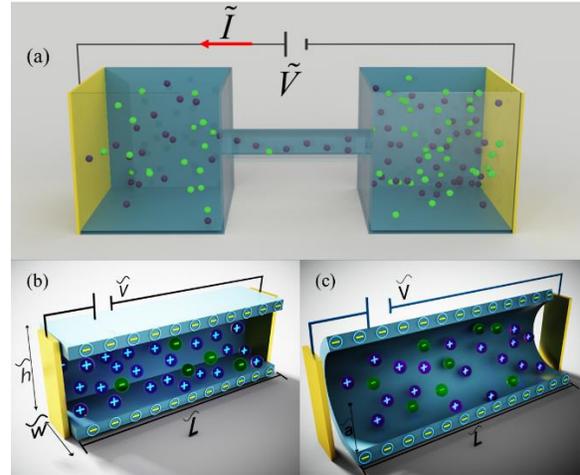

**Figure 1.** (a) Schematic of a realistic nanofluidic system, including the reservoirs with asymmetric salt concentrations, $c_{\text{left}}$ and $c_{\text{right}}$, at the two ends of the nanochannels. The entire system is subject to a voltage drop $V$ (defined as positive from left to right). (b-c) Schematics of two simplified nanochannel-only models with (b) 2D parallel plate and (c) cylindrical geometries. These two simplified models do not account for the electrical response effects of the reservoirs and entrance effects but do account for the concentration asymmetry. The negative surface charge density results in an excess of positive counterions, represented by purple spheres, over the negative coions, represented by green spheres. Both geometries have length $L$. The cylinder has a radius, $a$, while the parallel plate geometry has length, $L$, width $w$, and height $h$. Figure (c) is reproduced with permission from our previous work, Ref. [29], Copyright (2022) (American Physical Society).

* Email: yoavgreen@bgu.ac.il
ORCID numbers:      O.L.: 0000-0002-3415-7052
                              Y.G.: 0000-0002-0809-6575

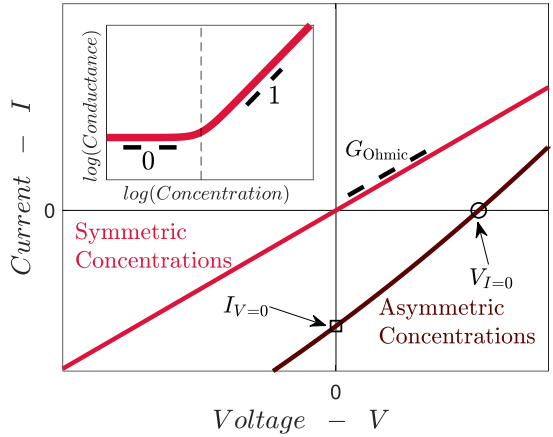

**Figure 2.** The current-voltage, $I-V$, response of symmetric concentrations and asymmetric concentrations systems. The slope is the linear Ohmic conductance, $G_{\mathrm{Ohmic}}$. When the concentrations are asymmetric, the $I-V$, which is not necessarily linear, is shifted. The two intercept points (the current at zero voltage, $I_{V=0}$, and the voltage at zero current, $V_{I=0}$) characterize the asymmetry. Inset: Schematic of the conductance, $G_{\mathrm{Ohmic}}$, versus the bulk concentration, $c_{\mathrm{bulk}}$. The slopes are denoted by the numbers below the dashed black lines.

**Electrical conductance of symmetric systems** ($c_{\mathrm{bulk}} = c_{\mathrm{left}} = c_{\mathrm{right}}$). The electrical conductance of a 'nanochannel' system, subjected to symmetric concentrations of binary (valences of $z_\pm = 1$) electrolyte with equal diffusion coefficients, $D_\pm = D$ such as KCl, is given by the well-known square root law [6,14,16,31] (**Figure 2** inset)

$$G_{\mathrm{Ohmic}}^{(\mathrm{symmetric})} = 2\frac{DF^2}{\Re T}\sqrt{\tfrac{1}{4}\Sigma_s^2 + c_{\mathrm{bulk}}^2}\frac{A}{L}. \qquad (1)$$

Here $L$ is the length of the channel and $A$ is the cross-section area, $\Re$ is the universal gas constant, $T$ is the absolute temperature, $F$ is the Faraday constant, and $c_{\mathrm{bulk}} = c_{\mathrm{left}} = c_{\mathrm{right}}$ is the bulk concentration on both sides of the nanochannel. Here, $\Sigma_s = -\sigma_s P/FA$, is the averaged excess counterion concentration (also known as the Dukhin number [32] – see Ref. [30] for more details), i.e., the average volumetric charge due to the surface charge density, $\sigma_s$, which is distributed along the perimeter, $P$, and divided by the cross-section area, $A$.

**Asymmetric concentrations** ($c_{\mathrm{left}} \neq c_{\mathrm{right}}$). Introduction of asymmetric bulk concentrations at the edges of the system naturally leads to the creation of ionic current from the reservoir of higher concentrations to the reservoir of lower concentrations. This is the current at zero voltage, $I_{V=0}$, and it is this current that can be harvested freely by RED systems. In order to nullify the current, an offset voltage is needed – this is the voltage at zero current, $V_{I=0}$.

To date, $V_{I=0}$ is known only in the ideal selectivity limit (where $\Sigma_s \gg c_{\mathrm{left,right}}$) and is given by $V_{I=0}^{(\mathrm{ideal\text{-}slectivity})} = V_{th}\ln(c_{\mathrm{right}}/c_{\mathrm{left}})$ with $V_{th} = \Re T/F$ being the thermal voltage. There are attempts to extend the expression of $V_{I=0}^{(\mathrm{ideal\text{-}slectivity})}$ to hold for all concentrations (i.e., all selectivities). However, these extensions are phenomenological. With regard to the zero-voltage current $I_{V=0}$, it is either measured experimentally or approximated by $I_{V=0} = V_{I=0}G_{I=0}$ where $G_{I=0}$ is the Ohmic resistance measured at zero current. This approximation is discussed further below.

**Novel model for asymmetric systems.** Using the Poisson-Nernst-Planck equations, we have derived an exact $I-V$ relation that provides expressions for $G_{\mathrm{Ohmic}}$, $V_{I=0}$, and $I_{V=0}$. In the main text, we will focus on the results and their outcomes, while the derivation of the model can be found in the Supplementary Information [30]. Also, in the SI [30], we describe the details of the numerical simulations that were used to verify the theoretical model. These simulations are shown in all subplots of **Figure 3** and show remarkable correspondence to our theory.

*Electrical conductance for asymmetric systems.* The Ohmic conductance (i.e., the conductance at low electric currents $I = 0$) for a system subject to asymmetric salt concentrations is given by

$$G_{\mathrm{Ohmic}} = 2\frac{DF^2}{\Re T}\frac{S_{\mathrm{left}} - S_{\mathrm{right}}}{\ln(S_{\mathrm{left}}/S_{\mathrm{right}})}\frac{A}{L}, \qquad (2)$$

where $S_{k=\mathrm{left,right}} = \sqrt{\tfrac{1}{4}\Sigma_s^2 + c_k^2}$.

Several observations regarding Eq. (2) are warranted. The most important observation is that in the limit of symmetric concentrations, $c_{\mathrm{left}} = c_{\mathrm{right}}\ (= c_{\mathrm{bulk}})$, $G_{\mathrm{Ohmic}}$ reduces to Eq. (1) such that $G_{\mathrm{Ohmic}} = G_{\mathrm{Ohmic}}^{(\mathrm{symmetric})} = 2DF^2 S_{\mathrm{bulk}}A/(\Re TL)$. The qualitative features of Eq. (2), shown in **Figure 3**(a), relative to $G_{\mathrm{Ohmic}}^{(\mathrm{symmetric})}$, remain virtually unchanged except for a few quantitative changes. At low concentration, at the limit of ideal selectivity, when $\Sigma_s \gg c_{\mathrm{left,right}}$, the response is still determined by $\Sigma_s$ such that $G_{\mathrm{Ohmic}}^{(\mathrm{ideal\text{-}slectivity})} = DF^2\Sigma_s A/(\Re TL)$. Next, we note that the transition from a bulk response to a surface charge-dominated response for a symmetric system occurs around $\Sigma_s \sim 2c_{\mathrm{bulk}}$. Naturally, this transition point varies for asymmetric concentrations and depends on the exact values of $c_{\mathrm{left}}$ and $c_{\mathrm{right}}$. As can be observed, the transition point strongly depends on the larger of the two values. The shift is more predominant when $c_{\mathrm{right}}/c_{\mathrm{left}} > 1$, since here $c_{\mathrm{right}}$ dominates the response. Finally, at very high concentrations ($c_{\mathrm{left,right}} \gg \Sigma_s$), we note that $G_{\mathrm{Ohmic}}^{(\mathrm{vanishing\text{-}slectivity})} = 2DF^2 A/(\Re TL)$ such that the conductance is linear with the concentration (i.e., a slope of 1). This behavior remains true also for Eq. (2) where the slope is linear with the concentrations, but the exact value depends on an interplay of the values of $c_{\mathrm{left}}$ and $c_{\mathrm{right}}$. This, too, can be expected since the conductance depends on the amount of available ions that can be transported.

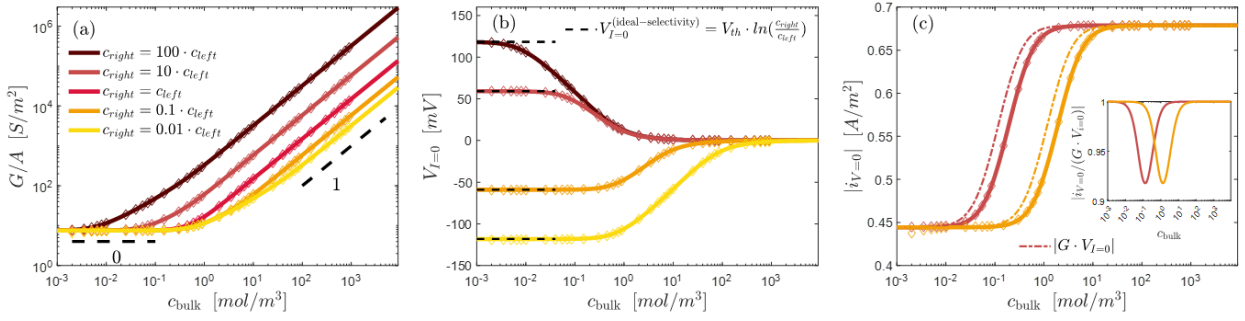

**Figure 3**. (a) The conductance density $G_{\text{Ohmic}}/A$, given by Eq. (2), versus the bulk concentration $c_{\text{bulk}} = c_{\text{left}}$ for several concentration gradients. (b) The zero-current voltage $V_{I=0}$, given by Eq. (3), versus the bulk concentration. The dashed black lines at low bulk concentrations represent the known limit of $V_{I=0}^{(\text{ideal-selective})}$. (c) The absolute value of the zero-voltage current density $i_{V=0}$ versus the bulk concentration. Solid lines represent values calculated from Eq. (4), while the dashed curves represent the approximation $|i_{V=0}| = |G_{\text{Ohmic}} V_{I=0}|$. The inset shows the ratio between the calculated values and the approximate values.

*Zero-current voltage*, $V_{I=0}$. The voltage at zero current is given by

$$V_{I=0} = V_{th}\left[\ln\left(\frac{c_{\text{right}}}{c_{\text{left}}}\right) + \ln\left(\frac{S_{\text{left}} + \tfrac{1}{2}\Sigma_s}{S_{\text{right}} + \tfrac{1}{2}\Sigma_s}\right)\right]. \quad (3)$$

We note that in the limit of $\Sigma_s \gg c_{\text{left,right}}$, the argument in the second term approaches unity such that the second term is negligible, and we recapitulate the known limit of ideal selectivity, $V_{I=0}^{(\text{ideal-slectivity})}$ [dashed black lines in **Figure 3**(b)]. Second, as can be expected, as the concentration is increased, where $\Sigma_s \ll c_{\text{left,right}}$, the permselectivity vanishes, and the voltage at zero current decreases until it reaches a value of zero. These two points can be observed in **Figure 3**(b).

*Zero-voltage current*, $I_{V=0}$. In contrast to $G_{\text{Ohmic}}$ and $V_{I=0}$ the current at zero voltage, $I_{V=0}$, is not given by a simple analytic expression. Instead, it is given by a transcendental equation for the current density, $i = I/A$ (note we use $I$ and $i$ interchangeably to denote the electric current )

$$i_{V=0}\ln\left(\frac{i_{V=0} - 2Fj_{V=0}S_{\text{right}}/\Sigma_s}{i_{V=0} - 2Fj_{V=0}S_{\text{left}}/\Sigma_s}\right) - Fj_{V=0}\frac{V_{I=0}}{V_{th}} = 0, \quad (4)$$

where $j_{V=0} = -D\Sigma_s V_{I=0}/(LV_{th}) + j_{I=0}$ is the salt current density at zero voltage, and it depends on the salt current density at zero current, $j_{I=0} = 2D(S_{\text{left}} - S_{\text{right}})/L$.

The excellent correspondence of the numerical solution of Eq. (4) [30] to exact numerical simulations, shown in **Figure 3**(c), confirms our results but also challenges the accepted paradigm on how to approximate $I_{V=0}$ and how to calculate the self-consistent power that can be harvested from RED systems.

In the simplest electrical system, comprised of a simple Ohmic resistor, the power is given by

$$P = IV = I^2/G_{\text{Ohmic}} = V^2 G_{\text{Ohmic}}. \quad (5)$$

It is important to note that in all scenarios, one utilizes $I, V$, and $G_{\text{Ohmic}}$ at the same conditions, which we shall now term as "local" quantities. Further, the two right terms in Eq. (5) hold so long Ohm's law holds (i.e., $V = IR_{\text{Ohmic}} = IG_{\text{Ohmic}}^{-1}$). As it turns out, this assumption doesn't universally hold for the simple nanofluidic system considered here, where the $I-V$ response is not purely linear. This is because the $I-V$ depends on an intermediate variable – the salt current $J$ – in a nonlinear manner (the detailed $I-V$ is provided in [30]).

In the RED literature, one often finds that the power is given by (here for clarity, we denote $G_{\text{Ohmic}} = G_{I=0}$)

$$P \approx I_{V=0}V_{I=0} \ (\text{or} \ P = I_{V=0}^2/G_{I=0} \ \text{or} \ P = V_{I=0}^2 G_{I=0}), \quad (6)$$

All these terms are equivalent and equally problematic, as they utilize quantities from two different states. For example, the first two expressions in Eq. (6) contain the implicit assumption that $I_{V=0} = V_{I=0}G_{I=0}$. The dashed lines in **Figure 3**(c) depict this incorrect approximation, where the lack of compatibility of the current for intermediate concentrations can be observed. This deviation is easily explained by previous observation that the $I-V$ is not linear (in contrast to simple conductors such as metals). In fact, upon further inspection of the $I-V$ shown in **Figure 2**, it can be discerned that the slopes (i.e., the <u>local</u> conductance) at the points $I_{V=0}$ and $V_{I=0}$ are different ($G_{I=0} \neq G_{V=0}$).

Naturally, if the current approximation is incorrect, so is the estimated power. The inset of **Figure 3**(c) inset demonstrates that the incorrect approximation overpredicts the current and thus overpredicts the power that can be harvested by the RED process by a few percent. Rather, the correct power $P = I_{V=0}^2/G_{V=0}$, depends on the local slope, $G_{V=0}$ (which can be calculated from the general $I-V$, see Ref. [30]).

In fact, from an experimental point of view, it is rather easy to calculate the power. One needs to scan the $I-V$ around $V = 0$, and then extract the local slope at this point. This is not a pure linear resistance, but rather it is the local differential resistance.

It is important to note that $P = V_{I=0}^2 G_{I=0}$ is not entirely wrong, as it uses two <u>local</u> quantities in the same state. First, there is an embedded use of Ohm's law even though the local current is zero. Second and more importantly, this power is not the relevant power for osmotic power generation.

**Conclusions and future directions.** This work considers the general current-voltage response of a nanofluidic system subject to a combined potential and concentration gradient. Here, we have derived an analytical model and presented novel expressions for $G_{\text{Ohmic}}$, $V_{I=0}$, and $I_{V=0}$ which are the key transport characteristics of RED systems.

Specifically, we have shown that $G_{\text{Ohmic}}$ [Eq. (2)] can be generalized for an arbitrary concentration gradient. The behavior of $G_{\text{Ohmic}}$ is consistent with expectations and finite element simulations. We also provide an expression for $V_{I=0}$ [Eq. (3)] that holds for all concentrations (and all selectivities). Finally, we present a transcendental equation [Eq. (4)] from which $I_{V=0}$ can be calculated. The calculated values differ from the accepted paradigm, and one outcome is that the maximal power that can be generated by RED needs to be calculated in a consistent manner.

We believe that this work can be extended to account for additional phenomena. Namely, the effects of the microchannels adjacent to the nanochannel, which gives rise to the access resistance, as well as the effects of surface charge regulation.

*Three layers.* Past works have shown that the contribution of the microchannels and entrance effects are not negligible, and they considerably change the response of the total conductance of the system [33–36]. We are already investigating how the microchannels vary $G_{\text{Ohmic}}$, $V_{I=0}$, and $I_{V=0}$. Our results will be published upon completion of our analysis.

*Surface charge regulation.* Here, we have explicitly assumed that $\Sigma_s$ (or $\sigma_s$) are concentration and spatially independent such that $\Sigma_s \sim c_{\text{bulk}}^0$. In recent years, it has been shown that the surface charge can be concentration-dependent through a surface charge regulation mechanism, leading to $\Sigma_s \sim c_{\text{bulk}}^\alpha$ (the exact value of $\alpha$ depends on a combination of the electrolyte and material properties) [6,29,37,38]. However, all these past works focused on symmetric concentrations ($c_{\text{left}} = c_{\text{right}}$) such that $\Sigma_s$ was still spatially independent. When $c_{\text{left}} \neq c_{\text{right}}$, $\Sigma_s$ will be spatially dependent, introducing additional challenges associated with "non-developed" flows. Thus, future works should consider surface charge regulation subject to asymmetric concentrations.

The results of this work provide a robust and simple tool for all scientists researching ion transport across permselective systems. Both experimentalists and MD numericists can use this model to design and calibrate their systems, while theoreticians can use this model to ensure that any newer model reproduces this model at the appropriate limits.

**Acknowledgments.** We thank Lyderic and Marie-Laure Bocquet, and their groups, for enlightening discussions during YG's visit to their labs. We thank Ramadan Abu-Rjal for his helpful feedback. We also thank Tomer Miara for his assistance with some of the illustrations. This work was supported by Israel Science Foundation grants 337/20 and 1953/20. We acknowledge the support of the Ilse Katz Institute for Nanoscale Science & Technology.